# Mining Attribute-structure Correlated Patterns in Large Attributed Graphs

Arlei Silva
Universidade Federal de Minas Gerais
Belo Horizonte, Brasil
arlei@dcc.ufmg.br

Wagner Meira Jr.
Universidade Federal de Minas Gerais
Belo Horizonte, Brasil
meira@dcc.ufmg.br

Mohammed J. Zaki
Rensselaer Polytechnic Institute
Troy, NY
zaki@cs.rpi.edu

## ABSTRACT

In this work, we study the correlation between attribute sets and the occurrence of dense subgraphs in large attributed graphs, a task we call structural correlation pattern mining. A structural correlation pattern is a dense subgraph induced by a particular attribute set. Existing methods are not able to extract relevant knowledge regarding how vertex attributes interact with dense subgraphs. Structural correlation pattern mining combines aspects of frequent itemset and quasi-clique mining problems. We propose statistical significance measures that compare the structural correlation of attribute sets against their expected values using null models. Moreover, we evaluate the interestingness of structural correlation patterns in terms of size and density. An efficient algorithm that combines search and pruning strategies in the identification of the most relevant structural correlation patterns is presented. We apply our method for the analysis of three real-world attributed graphs: a collaboration, a music, and a citation network, verifying that it provides valuable knowledge in a feasible time.

## 1. INTRODUCTION

In several real-life graphs, attributes can be associated with vertices in order to represent vertex properties. In social networks, for example, vertex attributes are useful to model personal characteristics. Moreover, vertex attributes can be associated with content (e.g., keywords, tags) in the web graph. Such an extended graph representation, which is called an attributed graph, may support graph patterns that provide relevant knowledge in various application scenarios.

An interesting question related to attributed graphs is how particular attributes are associated with the topology of real graphs. In other words, do there exist patterns that explain how vertex attributes interact with the graph structure? How can we extract and evaluate such patterns? In this paper, we study the problem of correlating attribute sets with an important topological property of graphs, which is the organization of vertices into dense subgraphs. For instance, we aim to address questions such as: How does a particular set of interests induce communities in a social network? What are the communities that emerge around such interests? Such questions are related to important social phenomena such as homophily [11] and influence [2]. Although several definitions of dense subgraphs have been proposed in the literature, most of them do not take vertex attributes into consideration. Furthermore, such definitions do not provide any knowledge regarding how different sets of attributes induce dense subgraphs.

This work studies the correlation between vertex attributes and dense subgraphs, a task we call *structural correlation pattern mining*. The structural correlation of an attribute set is the probability of a vertex to be member of a dense subgraph in its induced graph. Moreover, a structural correlation pattern is a dense subgraph induced by a particular attribute set. Figure 1 illustrates a dataset for structural correlation pattern mining. The vertex attributes are given in Figure 1(a) and the graph is shown in Figure 1(b). Example dense subgraphs are shown in Figures 1(c) and 1(d). The structural correlation of the attribute A is 0.82, since 9 out of 11 vertices are covered by dense subgraphs in its induced graph. On the other hand, the structural correlation of C is 0, because there is no dense subgraph inside the graph induced by C. The structural correlation of {A,B} is 1, due to the fact that every vertex is a member of a dense subgraph in the graph induced by {A,B}. The pair ({A,B}, {6,7,8,9,10,11}) is an example of a structural correlation pattern, for which the subgraph is shown in Figure 1(d). Another example is the pattern ({A}, {3,4,5,6}), for which the induced subgraph is shown in Figure 1(d).

The structural correlation of attribute sets and the structural correlation patterns are complementary information, while the first is a measure of the correlation between a given attribute set and the occurrence of dense subgraphs, the second provides representatives for such a correlation through specific subgraphs. We formulate the structural correlation pattern mining in terms of two existing data mining problems: frequent itemset and quasi-clique mining. Frequent itemset mining [1, 19] is applied to handle the possible large number of attribute sets from the graph and quasi-cliques [14, 10] are used as a definition for dense subgraphs.

We study structural correlation pattern mining focusing on two important aspects. The first aspect is the significance of the patterns. More specifically, it is relevant to provide significance measures for the structural correlation of attribute sets and the structural correlation patterns. The second aspect is related to the computational cost of the proposed task. Our objective is to enable the analysis of large real graphs in a feasible time. Although significance and high-performance are not necessarily concordant goals, we propose significance metrics that may lead to efficient pruning strategies for structural correlation pattern mining.

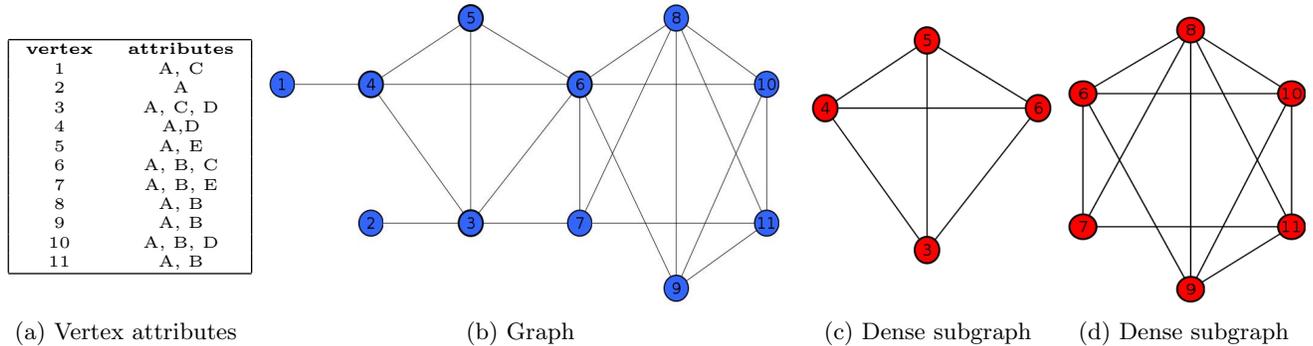

(a) Vertex attributes  (b) Graph  (c) Dense subgraph  (d) Dense subgraph

Figure 1: Structural correlation pattern mining (illustrative example)

Regarding the significance of patterns, we formulate normalization approaches for structural correlation pattern mining in order to measure the statistical significance of the structural correlation of a given attribute set. The idea is to compare the structural correlation against its expected value, which is provided by a null model. Moreover, we evaluate the structural correlation patterns in terms of size (i.e., number of vertices) and density (i.e., cohesion). Such evaluation is useful to rank the most interesting patterns.

We combine the statistical significance of the structural correlation of attribute sets and the size and density of structural correlation patterns with effective constraints to prune down the search space. Moreover, we propose two strategies for computing the structural correlation of attribute sets efficiently. These pruning and search techniques are integrated into the SCPM (Structural Correlation Pattern Mining) algorithm, which is described and evaluated in this paper. In particular, we apply SCPM to the analysis of three real attributed graphs: collaboration, music and citation networks. The results show that SCPM is able to extract relevant knowledge regarding how vertex attributes are correlated with dense subgraphs in large attributed graphs.

## 2. STRUCTURAL CORRELATION PATTERN MINING

### 2.1 Definitions

#### 2.1.1 Structural Correlation

We define an attributed graph as a 4-tuple $\mathcal{G} = (\mathcal{V}, \mathcal{E}, \mathcal{A}, \mathcal{F})$ where $\mathcal{V}$ is the set of vertices, $\mathcal{E}$ is the set of edges, $\mathcal{A} = \{a_1, a_2, \ldots a_n\}$ is the set of attributes, and $\mathcal{F} : \mathcal{V} \to P(\mathcal{A})$ is a function that returns the set of attributes of a vertex. $P$ is the power set function. Each vertex $v_i$ in $\mathcal{V}$ has a set of attributes $\mathcal{F}(v_i) = \{a_{i1}, a_{i2}, \ldots a_{ip}\}$, where $p = |\mathcal{F}(v_i)|$ and $\mathcal{F}(v_i) \subseteq \mathcal{A}$. Figure 1(b) shows an example of an attributed graph where the vertex attributes are given in Figure 1(a).

Given the set of attributes $\mathcal{A}$, we define an *attribute set $S$* as a subset of $\mathcal{A}$ ($S \subseteq \mathcal{A}$). Moreover, we denote by $\mathcal{V}(S) \subseteq \mathcal{V}$ the vertex set induced by $S$ (i.e., $\mathcal{V}(S) = \{v_i \in \mathcal{V} | S \subseteq \mathcal{F}(v_i)\}$) and by $\mathcal{E}(S) \subseteq \mathcal{E}$ the edge set induced by $S$ (i.e., $\mathcal{E}(S) = \{(v_i, v_j) \in \mathcal{E} | v_i, v_j \in \mathcal{V}(S)\}$). The graph $\mathcal{G}(S)$, induced by $S$, is the pair $(\mathcal{V}(S), \mathcal{E}(S))$. We also define a support function $\sigma$, which gives the number of occurrences of an attribute set in the input graph ($\sigma(S) = |\mathcal{V}(S)|$), i.e., the number of vertices that contain $S$.

The structural correlation function measures the correlation between a given attribute set and the occurrence of dense subgraphs in an attributed graph. We apply quasi-cliques as a definition for dense subgraphs. Quasi-cliques are a natural extension of the traditional clique definition.

DEFINITION 1. (***Quasi-clique***) *Given a minimum density threshold $\gamma_{min}$ ($0 < \gamma_{min} \leq 1$) and a minimum size threshold min_size, a quasi-clique is a maximal vertex set $Q$ such that for each $v \in Q$, the degree of $v$ in $Q$ is at least $\lceil \gamma_{min}.(|Q|-1) \rceil$ and $|Q| \geq min\_size$.*

Figures 1(c) and 1(d) are examples of an 1-quasi-clique of size 4 and a 0.6-quasi-clique of size 6, respectively, from the graph shown in Figure 1(b). The *quasi-clique mining problem* consists of identifying the quasi-cliques from a graph considering minimum size and density parameters, a problem known to be #P-hard [14, 17].

We define the structural correlation of an attribute set $S$ as the probability of a vertex $v$ with attribute $S$ to be part of a quasi-clique in $\mathcal{G}(S)$.

DEFINITION 2. (***Structural correlation function $\epsilon$***) *Given an attribute set $S$, the structural correlation of $S$, $\epsilon(S)$, is given as:*

$$\epsilon(S) = \frac{|\mathcal{K}_S|}{|\mathcal{V}(S)|} \quad (1)$$

*where $\mathcal{K}_S$ is the set of vertices in quasi-cliques in $\mathcal{G}(S)$.*

In the graph from Figure 1, $\mathcal{K}_{\{A\}} = \{3, 4, 5, 6, 7, 8, 9, 10, 11\}$, $\mathcal{K}_{\{C\}} = \{\}$ and $\mathcal{K}_{\{A,B\}} = \{6, 7, 8, 9, 10, 11\}$, and thus the corresponding values of $\epsilon(\{A\})$, $\epsilon(\{C\})$, and $\epsilon(\{A, B\})$ are 0.82, 0, and 1, respectively. Structural correlation measures the dependence between attribute set $S$ and the density of the associated vertices. It indicates how likely $S$ is to be part of dense subgraphs. Our formulation enables the identification of attributes that induce vertices that are well connected in the graph. In a social network, for instance, such attributes are of great interest since they may be related to homophily or influence. Nevertheless, it is also relevant to understand the dense subgraphs induced by attribute sets. We call *structural correlation pattern* a quasi-clique that is homogeneous w.r.t. an attribute set.

DEFINITION 3. *(Structural correlation pattern). A structural correlation pattern is a pair $(S,Q)$, where $S$ is an attribute set $(S \subseteq \mathcal{A})$, and $Q$ is a quasi-clique from the graph induced by $S$ $(Q \subseteq \mathcal{V}(S))$, given the quasi-clique parameters $\gamma_{min}$ and $min\_size$.*

The pair $(\{A\},\{3,4,5,6\})$ is an example of a size 4 structural correlation pattern with density 1 induced by the attribute A in the graph from Figure 1. Another example of a structural correlation pattern is $(\{A,B\},\{6,7,8,9,10,11\})$, which is a size 6 structural correlation pattern with density 0.6 induced by the attribute set $\{A,B\}$.

### 2.1.2 Structural Correlation Pattern Mining Problem

Based on the definition of structural correlation patterns and structural correlation function, we formulate the *structural correlation pattern mining problem*. It comprises the identification of the attribute sets correlated with dense subgraphs and the dense subgraphs induced by such attribute sets. We apply a minimum support threshold $\sigma_{min}$ for attribute sets in order to prune down the number of patterns.

DEFINITION 4. *(Structural correlation pattern mining problem). Given an attributed graph $\mathcal{G}(\mathcal{V},\mathcal{E},\mathcal{A},\mathcal{F})$, a minimum support threshold $\sigma_{min}$, a minimum quasi-clique density $\gamma_{min}$ and size $min\_size$, and a minimum structural correlation $\epsilon_{min}$, the structural correlation pattern mining consists of identifying the set of structural correlation patterns $(S,Q)$ from $\mathcal{G}$, such that $S$ is an attribute set for which $\sigma(S) \geq \sigma_{min}$, $\epsilon(S) \geq \epsilon_{min}$, and $Q$ is a $\gamma_{min}$-quasi-clique for which $Q \subseteq \mathcal{V}(S)$ and $|Q| \geq min\_size$.*

As an example, we consider the attributed graph shown in Figure 1 and the parameters $\sigma_{min}$, $\gamma_{min}$, $min\_size$ and $\epsilon_{min}$ set to 3, 0.6, 4, and 0.5, respectively. The set of structural correlation patterns are shown in Table 1. For each pattern, we give the pair (attribute set, dense subgraph), the respective quasi-clique size and density ($\gamma$), and the attribute set support ($\sigma$) and structural correlation ($\epsilon$).

| pattern | size | $\gamma$ | $\sigma$ | $\epsilon$ |
|---|---|---|---|---|
| $(\{A\},\{6,7,8,9,10,11\})$ | 6 | 0.60 | 11 | 0.82 |
| $(\{A\},\{3,4,5,6\})$ | 4 | 1 | 11 | 0.82 |
| $(\{A\},\{3,4,6,7\})$ | 4 | 0.67 | 11 | 0.82 |
| $(\{A\},\{3,5,6,7\})$ | 4 | 0.67 | 11 | 0.82 |
| $(\{A\},\{3,6,7,8\})$ | 4 | 0.67 | 11 | 0.82 |
| $(\{B\},\{6,7,8,9,10,11\})$ | 6 | 0.60 | 6 | 1.0 |
| $(\{A,B\},\{6,7,8,9,10,11\})$ | 6 | 0.60 | 6 | 1.0 |

**Table 1: Patterns from the graph shown in Figure 1**

Similar to the quasi-clique mining, the structural correlation pattern mining is #P-hard [17]. This is because the quasi-clique mining problem can be reduced to the structural correlation pattern mining by assigning the same attribute to each vertex from the graph and setting $\sigma_{min}$ to 1.

Structural correlation pattern mining is based on the structural correlation function, which measures how a given attribute set is associated with the occurrence of dense subgraphs in an attributed graph. However, it is important to assess the significance/interestingness of a given structural correlation, which is the subject of the next section.

### 2.1.3 Statistical Significance of the Structural Correlation

Given the structural correlation of an attribute set, how can we evaluate it? In other words, what can be considered a high or low structural correlation? In this section, we address such questions by proposing null models for structural correlation. These models specify the expected structural correlation of an attribute set assuming that the correlation between vertex attributes and dense subgraphs is random. *Normalized structural correlation* measures how the structural correlation of an attribute set deviates from its expected value, and allows us to assess the statistical significance of a given structural correlation value.

DEFINITION 5. *(Normalized structural correlation). Given an attribute set $S$ with support $\sigma(S)$ and a function $\epsilon_{exp}$, which gives the expected structural correlation of an attribute set based on its support and the attributed graph $\mathcal{G}$, the normalized structural correlation of $S$ is given by:*

$$\delta(S,\mathcal{G}) = \frac{\epsilon(S)}{\epsilon_{exp}(\sigma(S),\mathcal{G})} \quad (2)$$

According to Definition 5, the normalized structural correlation function gives how much the structural correlation of an attribute set $S$ is higher than expected. Therefore, it requires the definition of the function $\epsilon_{exp}$, which receives the support of $S$ ($\sigma(S)$) and the attributed graph $\mathcal{G}$ as arguments. By normalizing the structural correlation, we expect to obtain a measure of the correlation of an attribute set $S$ that is independent of its support and the input graph.

We assume that the input graph $\mathcal{G}$ comprises the object of interest, i.e., it is the "population" graph. Assume that we are given the attribute set support value $\sigma(S)$ (independent of the actual attribute set $S$). To compute the expected structural correlation, our sample space is the set of all vertex subsets of size $\sigma(S)$ drawn randomly from $\mathcal{G}$. The statistic of interest is the mean structural correlation value, $\epsilon_{exp}$. That is, the expected probability that a random vertex in a given sample induces dense subgraphs (quasi-cliques) in that sample of size $\sigma(S)$. The quasi-clique parameters, $\gamma_{min}$ and $min\_size$, are assumed to be fixed as well.

An intuitive approach for computing $\epsilon_{exp}$ is through simulation. Given the support $\sigma(S)$ of the attribute set, a random sample of $\sigma(S)$ vertices from $\mathcal{G}$ is selected. Each vertex from the sample is checked to be in a quasi-clique, according to the quasi-clique parameters. The structural correlation of the sample is the fraction of vertices from it that are in at least one quasi-clique. The simulation-based expected structural correlation $sim$-$\epsilon_{exp}$ is given by the average structural correlation of $r$ random samples.

The simulation-based structural correlation is very simple conceptually but may require a high $r$ to achieve accurate estimates, which is prohibitive in real settings. Thus we also propose an analytical formulation for an upper bound on the expected structural correlation of an attribute set. The idea is that a vertex must have a minimum degree of $\lceil \gamma_{min}.(min\_size - 1) \rceil$ in order to be member of a $\gamma_{min}$-quasi-clique of minimum size $min\_size$. Consequently, the probability of a vertex to have a degree of $\lceil \gamma_{min}.(min\_size-1) \rceil$ in a random subgraph of size $\sigma(S)$ from $\mathcal{G}$ gives an upper bound on the expected structural correlation of $S$.

Given a random size $\sigma(S)$ subgraph $\mathcal{G}_{\sigma(S)}$ from $\mathcal{G}$, the degree of $v$ in $\mathcal{G}$ and $\mathcal{G}_{\sigma(S)}$ are related as follows.

THEOREM 1. *(Probability of a vertex that has a degree $\alpha$ in $\mathcal{G}$ to have a degree $\beta$ in $\mathcal{G}_{\sigma(S)}$).* If a random vertex $v$ from $\mathcal{G}$ with degree $\alpha$ is selected to be part of $\mathcal{G}_{\sigma(S)}$, the probability of such vertex to have a degree $\beta$ in $\mathcal{G}_{\sigma(S)}$ is given by the following binomial function:

$$F(\alpha, \beta, \rho) = \binom{\alpha}{\beta}.\rho^\beta.(1-\rho)^{\alpha-\beta} \quad (3)$$

where $\rho$ is the probability of a specific vertex $u$ from $\mathcal{G}$ to be in $\mathcal{G}_{\sigma(S)}$, if $v$ is already chosen, which is given as:

$$\rho = \frac{\sigma(S) - 1}{|\mathcal{V}| - 1} \quad (4)$$

**Proof sketch.** There are $\alpha$ vertices adjacent to $v$ in $\mathcal{G}$, thus, the probability of $v$ to have a degree of $\beta$ in $\mathcal{G}_{\sigma(S)}$ is the probability of selecting $\beta$ out of $\alpha$ vertices to be part of $\mathcal{G}_{\sigma(S)}$. Since $v$ is already selected, the probability of selecting any remaining vertex from $\mathcal{G}$ is given by equation 4.

Based on Theorem 1, we define an upper bound on the expected structural correlation as the probability of a vertex to have a degree of at least $\lceil \gamma_{min}.(min\_size - 1) \rceil$ in $\mathcal{G}_{\sigma(S)}$.

THEOREM 2. *(Upper bound on the expected structural correlation).* Given the quasi-clique parameters $\gamma_{min}$ and $min\_size$, the structural correlation of an attribute set with support $\sigma(S)$ is upper bounded by:

$$max\text{-}\epsilon_{exp}(\sigma(S)) = \sum_{\alpha=z}^{m} p(\alpha). \sum_{\beta=z}^{\alpha} F(\alpha, \beta, \rho) \quad (5)$$

where $z = \lceil \gamma_{min}.(min\_size - 1) \rceil$, $m$ is the maximum degree of a vertex from $\mathcal{G}$, and $p$ is the degree distribution of $\mathcal{G}$.
**Proof sketch.** Given a vertex with degree $\alpha$ in $\mathcal{G}$, the probability of such vertex to have a degree of at least $\lceil \gamma_{min}.(min\_size - 1) \rceil$ in $\mathcal{G}_{\sigma(S)}$ is the sum of expression 3 over the degree interval from $\lceil \gamma_{min}.(min\_size - 1) \rceil$ to $\alpha$. If we multiply this sum by the probability of a vertex of degree $\alpha$ from $\mathcal{G}$ to be in $\mathcal{G}_{\sigma(S)}$, i.e., $p(\alpha)$, it gives the probability of any vertex with degree $\alpha$ from $\mathcal{G}$ to have a degree of at least $\lceil \gamma_{min}.(min\_size - 1) \rceil$ in $\mathcal{G}_{\sigma(S)}$. Equation 5 is the sum of such products over the vertex degrees higher than $\lceil \gamma_{min}.(min\_size - 1) \rceil$.

The proposed upper bound on the expected structural correlation of an attribute $S$ is based on the expected degree distribution of a random graph of size $\sigma(S)$ from $\mathcal{G}$. However, the degree is not the only criteria for a vertex to be part of a quasi-clique. Vertices that satisfy the minimum degree threshold may not be part of a quasi-clique if they are connected to low degree vertices. Nevertheless, since we apply the proposed formulation in order to normalize the structural correlation of attribute sets with different supports, our objective is to provide a function that presents a slope that is similar to expected structural correlation. In Section 4.1, we compare the expected structural correlation computed using simulation with the proposed upper bound.

We call $\delta_{sim}$ and $\delta_{lb}$ the normalized structural correlation functions that apply the expected structural correlation based on simulation $sim\text{-}\epsilon_{exp}$ and the theoretical upper bound $max\text{-}\epsilon_{exp}$, respectively. Since $max\text{-}\epsilon_{exp} \geq sim\text{-}\epsilon_{exp}$, $\delta_{lb} = \frac{\epsilon(S)}{max\text{-}\epsilon_{exp}} \leq \frac{\epsilon(S)}{sim\text{-}\epsilon_{exp}} = \delta_{sim}$, thus, $\delta_{lb}$ is a lower bound on $\delta_{sim}$.

It is important to notice that $max\text{-}\epsilon_{exp}$ is monotonically non-decreasing, i.e., $max\text{-}\epsilon_{exp}(\sigma_1) > max\text{-}\epsilon_{exp}(\sigma_2)$ if and only if $\sigma_1 \geq \sigma_2$. It follows directly from the fact that the analytical upper bound (Equation 5) is based on a cumulative binomial function, which is known to be monotonically non-decreasing w.r.t. $\rho$. We also assume that $sim\text{-}\epsilon_{exp}$ is monotonically non-decreasing for sufficiently high values of $r$, since an increase in the size of the random graphs selected from $\mathcal{G}$ is not expected to decrease the probability of finding a vertex in a quasi-clique. Such properties will be exploited by our pruning techniques, which will be proposed further in this paper (see Section 3.2.1).

We apply the normalized structural correlation in the identification of statistically significant structural correlation values. Therefore, we extend the structural correlation pattern mining problem (Definition 4) by adding a minimum normalized structural correlation threshold $\delta_{min}$. Such a threshold may also be useful to improve the performance of structural correlation pattern mining algorithms, as will be discussed in Section 3.2. Since a user may be interested in patterns that have high structural correlation ($\epsilon$) as well as being statistically significant ($\delta$), we present results using both regular and normalized structural correlation.

## 2.2 Related Work

Finding communities [6, 3] and dense subgraphs [5, 10, 8, 20] has been an active research topic. A community is usually defined as set of vertices significantly more connected among themselves than with vertices outside it [3]. On the other hand, dense subgraphs, such as cliques [18], are strongly based on internal cohesion and maximality.

This work applies a dense subgraph definition called quasi-clique, which is a set of vertices where each vertex is connected at least to a fraction of the others. [14] introduces the problem of mining cross-graph quasi-cliques. They further studied the problem of mining frequent cross-graph quasi-cliques [8]. In [20] and [21] the authors study the problem of mining frequent coherent closed quasi-cliques. [10] studies the problem of finding quasi-cliques from a single graph, proposing pruning techniques for quasi-clique mining.

Graph clustering and dense subgraph discovery methods that consider vertex attributes as complementary information have attracted the interest of the research community in the recent years [12, 4, 22, 13]. A general assumption of these methods is that clusters based on both the topology of the graph and the attributes of vertices are more meaningful than those based only on the topology or the attributes. [4] proposes two efficient algorithms for the connected k-center problem, which has as objective to partition a graph considering both the attributes and the topology. [22] proposes a random walk-based distance metric in an augmented graph where vertices from the original graph are connected to new vertices that represent vertex attributes. In [12], the authors introduce the problem of mining cohesive patterns, which are dense connected subgraphs where vertices have homogeneous attributes (or features). [13] considers the problem of computing maximal homogeneous cliques in attributed graphs. Different from these methods, structural correlation pattern mining does not assume that vertex attributes are complementary information. In fact, we are interested in finding attribute sets that explain the formation of dense subgraphs through correlation.

Assessing how vertex attributes are related to the graph topology has led to the definition of new patterns. [15] proposed the problem of finding itemset-sharing subgraphs, which consists of extracting subgraphs with common itemsets. It is important to notice that such method do not consider the density of subgraphs. [9] defines the proximity pattern mining, which evaluates how close vertex attributes are in the graph. A proximity pattern is a set of labels that co-occur in neighborhoods. Therefore, proximity patterns are not necessarily dense subgraphs or cohesive, differently from structural correlation patterns. In [7], the authors propose a different definition for the structural correlation, which compares the closeness among vertices induced by a given single attribute against a subgraph where attributes are randomly distributed. Our work differs from [7] by combining multiple attributes and considering a particular topological property which is the organization into dense subgraphs. Moreover, besides the evaluation of structural correlation of attribute sets, we are interested in the discovery of relevant dense subgraphs to be representatives of the structural correlation.

In [16], we introduce the structural correlation pattern mining and present an algorithm for this problem called SCORP. In this paper, we study the problem of identifying statistically significant structural correlation patterns based on a normalization of the structural correlation. We also present the SCPM algorithm, which extends SCORP with new pruning and search strategies for structural correlation pattern mining. Different from SCORP, SCPM enumerates the top structural correlation patterns in terms of size and density efficiently, instead of the complete set of patterns.

## 3. ALGORITHMS

### 3.1 Naive Algorithm

Since structural correlation pattern mining combines aspects of the frequent itemset mining and the quasi-clique mining problems, we may combine a frequent itemset mining algorithm and a quasi-clique mining algorithm into a naive algorithm for structural correlation pattern mining.

The naive algorithm solves the structural correlation pattern mining problem (see Definition 4) by first enumerating the set of frequent attribute sets $\mathcal{F}$ from $\mathcal{G}$ and then identifying the set of quasi-cliques $\mathcal{Q}$ from the graph induced by each frequent attribute set $S$ from $\mathcal{F}$. The structural correlation of each frequent attribute set $S$ is computed by checking whether each vertex $v \in \mathcal{V}(S)$ is part of a quasi-clique in $\mathcal{Q}$. Frequent attribute sets can be identified using a frequent itemset mining algorithm [1, 19]. In this work, we apply the Eclat algorithm [19]. Moreover, any algorithm for quasi-clique mining can be applied by such naive algorithm. We apply the Quick algorithm [10].

The main drawback of the naive algorithm is that it enumerates the complete set of frequent attribute sets from $\mathcal{G}$ and the complete set of quasi-cliques from each induced graph $\mathcal{G}(S)$, where $S$ is a frequent attribute set. Since the frequent itemset mining and the quasi-clique mining problems are known to be #P-hard, the naive algorithm is expected to not be able to process large attributed graphs. In order to achieve such goal, in the upcoming sections, we describe several strategies for efficient structural correlation pattern mining. We combine such strategies into a new algorithm, which is described in Section 3.2. Further in this paper, we compare the performance of the proposed algorithm against this naive method.

### 3.2 SCPM Algorithm

This section presents the SCPM (Structural Correlation Pattern Mining) algorithm, which applies several strategies in order to enable the structural correlation pattern mining in large attributed graphs. Unlike the naive algorithm, SCPM does not enumerate every frequent attribute set but prunes those attribute sets that cannot satisfy a minimum structural correlation threshold. Moreover, instead of identifying each quasi-clique from an induced graph, SCPM checks whether vertices are in quasi-cliques by verifying a reduced number of quasi-clique candidates. Finally, SCPM returns the set of the top-k most relevant structural correlation patterns from the attributed graph.

#### 3.2.1 Pruning Strategies for SCP Mining

This section presents pruning techniques for structural correlation pattern mining. The objective of these pruning techniques is to reduce the execution time of the structural correlation pattern mining algorithms without compromising its correctness. Theorem 3 allows the pruning of vertices during the level-wise enumeration of attribute sets.

THEOREM 3. *(**Vertex pruning for attribute sets**). Let $\mathcal{K}_S$ be the set of vertices in dense subgraphs in the graph induced by an attribute set $S$. If $S_i \subseteq S_j$, then $\mathcal{K}_{S_j} \subseteq \mathcal{K}_{S_i}$.*
**Proof sketch.** *Lets suppose that there exists a vertex $v$ such that $v \in \mathcal{K}_{S_j}$ and $v \notin \mathcal{K}_{S_i}$. Since $v \in \mathcal{K}_{S_j}$, there exists a dense subgraph $V \subseteq \mathcal{V}(S_j)$, such that $v \in V$. Moreover, if $v \notin \mathcal{K}_{S_i}$, there does not exist any dense subgraph $U \subseteq \mathcal{V}(S_i)$ such that $v \in U$. Nevertheless, if $S_i \subseteq S_j$, then $\mathcal{V}(S_j) \subseteq \mathcal{V}(S_i)$, which implies that $V \subseteq \mathcal{V}(S_i)$ (contradiction).*

Based on Theorem 3, we can prune vertices that are not in dense subgraphs in the graph induced by a given attribute set before extending it to generate larger attribute sets. Attribute sets can also be pruned based on an upper bound on the structural correlation function, as stated by Theorem 4.

THEOREM 4. *(**Attribute set pruning based on the upper bound on the structural correlation**). For two attribute sets $S_i$ and $S_j$, if $S_i \subseteq S_j$ and $\sigma(S_j) \geq \sigma_{min}$, then $\epsilon(S_j) \leq \epsilon(S_i).|\mathcal{V}(S_i)|/\sigma_{min}$*
**Proof sketch.** *According to Theorem 3, $\epsilon(S_i).|\mathcal{V}(S_i)| \geq \epsilon(S_j).|\mathcal{V}(S_j)|$, since every vertex covered by a dense subgraph in $\mathcal{V}(S_j)$ is also covered by a dense subgraph in $\mathcal{V}(S_i)$. Moreover, since $\sigma(S_j) \geq \sigma_{min}$, $\epsilon(S_j)$ is upper bounded by $\epsilon(S_i).|\mathcal{V}(S_i)|/\sigma_{min}$ based on the definition of the structural correlation function $\epsilon$ (see Definition 2).*

Given an attribute set $S_i$, of size $i$, if $\epsilon(S_i).|\mathcal{V}(S_i)|/\sigma_{min} < \epsilon_{min}$, then $S_i$ is not included in the set of attribute sets to be combined for the generation of size $i + 1$ attribute sets. Theorem 4 guarantees that there does not exist an attribute set $S_j$, such that $S_i \subseteq S_j$ and $\epsilon(S_j) \geq \epsilon_{min}$. A similar pruning rule can be formulated based on the normalized structural correlation function definition.

THEOREM 5. *(**Attribute set pruning based on the upper bound on the normalized structural correlation**). For two attribute sets $S_i$ and $S_j$, if $S_i \subseteq S_j$, $\epsilon_{exp}$ is a monotonically non-decreasing, and $\sigma(S_j) \geq \sigma_{min}$, then $\delta(S_j) \leq \epsilon(S_i).|\mathcal{V}(S_i)|/(\epsilon_{exp}(\sigma_{min}).\sigma_{min})$*
**Proof sketch.** *According to Theorem 4, $\epsilon(S_j) \leq \epsilon(S_i).|\mathcal{V}(S_i)|/\sigma_{min}$. Since $\sigma(S_j) \geq \sigma_{min}$ and $\epsilon_{exp}$ is*

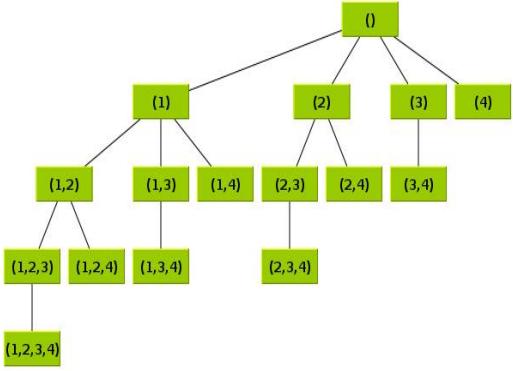

**Figure 2: Set enumeration tree**

**Algorithm 1** General Structural Correlation Algorithm

**Require:** $\mathcal{G}(S)$, $\gamma_{min}$, $min\_size$
**Ensure:** $\mathcal{Q}$
1: $\mathcal{Q} \leftarrow \emptyset$
2: $X \leftarrow \emptyset$
3: $candExts(X) \leftarrow \mathcal{V}(S)$
4: Apply vertex pruning in $candExts(X)$
5: $qcCands \leftarrow \{(X, candExts(X))\}$
6: **while** $qcCands \neq \emptyset$ **do**
7: $\quad q \leftarrow qcCands.get()$
8: $\quad$ Apply candidate quasi-clique pruning in $q$
9: $\quad$ **if** $q.X \cup q.candExts(X)$ is a quasi-clique **then**
10: $\quad\quad \mathcal{Q} \leftarrow \mathcal{Q} \cup \{q.X \cup q.candExts(X)\}$
11: $\quad$ **else**
12: $\quad\quad$ **if** $q.X$ is a quasi-clique **then**
13: $\quad\quad\quad \mathcal{Q} \leftarrow \mathcal{Q} \cup \{q.X\}$
14: $\quad\quad$ **end if**
15: $\quad\quad$ insert extensions of $q$ into $qcCands$
16: $\quad$ **end if**
17: **end while**

monotonically non-decreasing, then $\epsilon_{exp}(\sigma(S_j)) \geq \epsilon_{exp}(\sigma_{min})$. Therefore, $\delta(S_j) \leq \epsilon(S_i).|\mathcal{V}(S_i)|/(\epsilon_{exp}(\sigma_{min}).\sigma_{min})$.

If $\delta(S_i).|\mathcal{V}(S_i)|/(\epsilon_{exp}(\sigma_{min}).\sigma_{min}) < \delta_{min}$, the attribute set $S_i$, of size $i$, is not included in the set of attribute sets to be combined for the generation of size $i + 1$ attribute sets. Since $\delta_{lb}$ gives a lower bound on the normalized structural correlation, the whole pruning potential of Theorem 5 may not be explored. Nevertheless, the results show that use of $\delta_{lb}$ enables significant performance gains (see Section 4.2).

The pruning strategy stated by Theorem 3 reduces the number of vertices to be checked to be in quasi-cliques in the computation of structural correlation. Theorems 4 and 5 enable the reduction of the attribute sets for which the structural correlation is computed to a set that is expected to be smaller than the set of frequent attribute sets.

### 3.2.2 Computing the Structural Correlation

As discussed in Section 3.1, the naive algorithm computes the structural correlation of an attribute set $S$ through the enumeration of the quasi-cliques from $\mathcal{G}(S)$. In this section, we describe how the structural correlation can be computed by identifying a reduced number of quasi-clique candidates.

Quasi-cliques can be enumerated based on a vertex set $X$, initially set as $\emptyset$, and a set of candidate extensions of $X$, $candExts(X)$, initially set as $\mathcal{V}$. Vertices are moved from $candExts(X)$ to $X$, one at a time, until the complete set of quasi-clique candidates are generated. Figure 2 shows a set enumeration tree that represents the search space of quasi-cliques considering a set of 4 vertices (1-4). In order to prune down such search space, quasi-clique mining algorithms apply several pruning techniques. We divide these techniques into two groups:

1. **Vertex pruning:** Removal of vertices that cannot be part of any quasi-clique in $\mathcal{G}$ according to the quasi-clique definition and the quasi-clique parameters. Vertex pruning is performed iteratively over the graph in order to minimize the search space of quasi-cliques.

2. **Candidate quasi-clique pruning:** Removal of candidate quasi-cliques (i.e., pairs $(X, candExts(X))$) from the search space of quasi-cliques. Such removal is based on the properties of the subgraph composed by vertices from $X$ and $candExts(X)$.

Algorithm 1 gives a general description of how quasi-cliques are identified in the computation of structural correlation. This algorithm is also used as the basis for the enumeration of the top-k structural correlation patterns. The algorithm receives an induced graph $\mathcal{G}(S)$, and the minimum density ($\gamma_{min}$) and size ($min\_size$) for quasi-cliques. It gives as output a set of quasi-cliques $\mathcal{Q}$ from $\mathcal{G}$. Vertex and quasi-clique candidate prunings are applied in lines 4 and 8, respectively. Candidate quasi-cliques are managed by the data structure $qcCands$, which will be discussed later. Each candidate pattern is checked to be a lookahead quasi-clique (i.e., $q.X \cup q.candExts(X)$ is a quasi-clique) first, due to the fact that quasi-cliques are maximal. In case such a condition does not hold, $q.X$ is checked to be a quasi-clique and the extensions of $q$ are inserted into $qcCands$ (line 15). The algorithm finishes when $qcCands$ becomes empty. The set $\mathcal{K}_S$, which is composed of vertices covered by quasi-cliques in $\mathcal{G}(S)$, can be obtained directly from $\mathcal{Q}$.

Since the quasi-clique mining problem is known to be #P-hard, the identification of quasi-cliques may require processing a large number of quasi-clique candidates, which would constitute an important limitation to the computation of the structural correlation of large induced graphs. Nevertheless, computing the structural correlation does not require the enumeration of the complete set of quasi-cliques. The necessary information is whether each vertex from the induced graph is covered by a quasi-clique or not. Therefore, candidate quasi-cliques composed of vertices already known to be covered by quasi-cliques can be pruned from the new quasi-clique candidates generated in line 15 of Algorithm 1.

Besides pruning candidate quasi-cliques that are already known to be covered by dense subgraphs, we also propose search strategies for computing the structural correlation. These search strategies determine the order in which candidate quasi-cliques are enumerated. A breadth-first search (BFS) strategy for computing the structural correlation traverses the search space of quasi-cliques in a breadth-first order, starting from the root and visiting the smaller vertex sets before the larger ones. On the other hand, a depth-first search (DFS) strategy extends vertex sets as much as possible. The BFS strategy is expected to perform better in case covering vertices with smaller quasi-cliques is more efficient than with larger quasi-cliques. Considering a set of 4 vertices, for which the search space of quasi-cliques is shown in Figure 2, the BFS and the DFS strategy visit the quasi-clique candidates as follows:

**Algorithm 2** SCPM Algorithm

**Require:** $\mathcal{G}, \sigma_{min}, \gamma_{min}, min\_size, \epsilon_{min}, \delta_{min}, k$
**Ensure:** $\mathcal{P}$
1: $\mathcal{P} \leftarrow \emptyset$
2: $\mathcal{T} \leftarrow \emptyset$
3: $\mathcal{I} \leftarrow$ frequent attributes from $\mathcal{G}$
4: **for all** $S \in \mathcal{I}$ **do**
5:   $\epsilon \leftarrow$ structural correlation of $S$
6:   **if** $\epsilon \geq \epsilon_{min}$ AND $\epsilon/\epsilon_{exp}(S) \geq \delta_{min}$ **then**
7:     $\mathcal{Q} \leftarrow$ top-k patterns from $\mathcal{G}(S)$
8:     **for all** $q \in \mathcal{Q}$ **do**
9:       $\mathcal{P} \leftarrow \mathcal{P} \cup (S, q)$
10:     **end for**
11:   **end if**
12:   **if** $\epsilon.\sigma(S) \geq \epsilon_{min}.\sigma_{min}$ AND $\epsilon.\sigma(S) \geq \delta_{min}.\epsilon_{exp}(\sigma_{min}).\sigma_{min}$ **then**
13:     $\mathcal{T} \leftarrow \mathcal{T} \cup S$
14:   **end if**
15: **end for**
16: $\mathcal{P} \leftarrow \mathcal{P} \cup$ enumerate-patterns$(\mathcal{T}, \mathcal{G}, \sigma_{min}, \gamma_{min}, min\_size, \epsilon_{min}, \delta_{min}, k)$

**Algorithm 3** enumerate-patterns

**Require:** $\mathcal{T}, \mathcal{G}, \sigma_{min}, \gamma_{min}, min\_size, \epsilon_{min}, \delta_{min}, k$
**Ensure:** $\mathcal{P}$
1: $\mathcal{P} \leftarrow \emptyset$
2: **for all** $S_i \in \mathcal{T}$ **do**
3:   $\mathcal{R} \leftarrow \emptyset$
4:   **for all** $S_j \in \mathcal{T}$ **do**
5:     **if** $i > j$ **then**
6:       $S \leftarrow S_i \cup S_j$
7:       **if** $\sigma(S) \geq \sigma_{min}$ **then**
8:         $\epsilon \leftarrow$ structural correlation of $S$
9:         **if** $\epsilon \geq \epsilon_{min}$ AND $\epsilon/\epsilon_{exp}(S) \geq \delta_{min}$ **then**
10:           $\mathcal{Q} \leftarrow$ top-k patterns from $\mathcal{G}(S)$
11:           **for all** $q \in \mathcal{Q}$ **do**
12:             $\mathcal{P} \leftarrow \mathcal{P} \cup (S, q)$
13:           **end for**
14:         **end if**
15:         **if** $\epsilon.\sigma(S) \geq \epsilon_{min}.\sigma_{min}$ AND $\epsilon.\sigma(S) \geq \delta_{min}.\epsilon_{exp}(\sigma_{min}).\sigma_{min}$ **then**
16:           $\mathcal{R} \leftarrow \mathcal{R} \cup S$
17:         **end if**
18:       **end if**
19:     **end if**
20:   **end for**
21:   $\mathcal{P} \leftarrow \mathcal{P} \cup$ enumerate-patterns$(\mathcal{R}, \mathcal{G}, \sigma_{min}, \gamma_{min}, min\_size, \epsilon_{min}, \delta_{min}, k)$
22: **end for**

- **BFS:** $\{1\}, \{2\}, \{3\}, \{4\}, \{1,2\}, \ldots \{1,2,3,4\}$.
- **DFS:** $\{1\}, \{1,2\}, \{1,2,3\}, \{1,2,3,4\}, \{1,3\}, \ldots \{4\}$.

Quasi-cliques can be enumerated in BFS order by using a queue as a data structure to manage quasi-clique candidates in Algorithm 1. Similarly, a DFS strategy for enumerating quasi-cliques can apply a stack in order to manipulate candidate patterns. Further in this paper, we evaluate the search strategies presented in this section.

### 3.2.3 Enumerating Top-k Patterns

As discussed in Section 2.1.2, enumerating structural correlation patterns is a computationally expensive task. In this section, we study how to reduce the cost of enumerating structural correlation patterns by restricting the output set to only the top-k most relevant patterns in terms of size (primary criteria) and density (secondary criteria).

The enumeration of the top-k structural correlation patterns follows the same procedure described in Algorithm 1. We use a DFS strategy in the discovery of the top-k patterns because structural correlation patterns are maximal (see Definition 3). However, since the number of patterns to be discovered is known, a current set of patterns can be applied to prune the search space of new candidates. New candidate quasi-cliques are generated in line 15. In case the current set of top patterns contains k patterns and a candidate pattern $p$ cannot produce a pattern larger than the smallest current top-k pattern $t$ (i.e., $|p.X \cup p.candExts(X)| < |t|$), $p$ can be pruned. By updating the set of top-k patterns, the minimum size threshold is increased iteratively. As a consequence, the top-k patterns are enumerated more efficiently than the complete set of patterns from an induced graph.

Algorithm 2 is a high-level description of the SCPM algorithm, which applies the strategies for efficient structural correlation pattern mining presented in this section. The initial set of attributes $\mathcal{I}$ is composed by those with a support of at least $\sigma_{min}$ (line 3). The structural correlation of each size one attribute set $S \in \mathcal{I}$ is computed as described in Section 3.2.2. In case the structural correlation of $S$ satisfies minimum structural correlation ($\epsilon_{min}$) and normalized structural correlation ($\delta_{min}$) thresholds, the top-k patterns induced by $S$ are identified using the algorithm described in this section (line 7). These patterns are included into a set of patterns $\mathcal{P}$ that will be given as output. The pruning rules for attribute sets based on $\epsilon$ and $\delta$ (see Section 3.2.1) are applied in line 12. Pruned attributes are not included into the set of attributes $\mathcal{T}$ to be extended. These attributes are extended by the function enumerate-patterns (line 16).

Algorithm 3 describes the function enumerate-patterns. It receives the same input parameters of SCPM, and also the set of patterns to be extended $\mathcal{T}$. It returns the set of top-k patterns $(S, V)$ that have attribute sets extended from those in $\mathcal{T}$ regarding the input parameters. New attribute sets are extended through the union of existing ones (line 6). Attribute sets are traversed in a DFS order (e.g., $\{A\}, \{A, B\}, \{A, B, C\} \ldots \{E\}$). The enumerate-patterns function is similar to Algorithm 2, except that each new attribute set $S$ is checked to satisfy the minimum support threshold $\sigma_{min}$ (line 7). All valid attribute sets are generated through recursive calls to enumerate-patterns (line 21).

## 4. EXPERIMENTAL RESULTS

This section presents case studies on the structural correlation pattern mining using real datasets. Moreover, we evaluate the performance and study the sensitivity of important input parameters of SCPM. Experiments were executed on a 16-core Intel Xeon 2.4 Ghz with 50GB of RAM. The implementations are available as open-source[1].

### 4.1 Case Studies

#### 4.1.1 DBLP

In the attributed graph extracted from the DBLP[2] digital library, each vertex represents an author and two authors are connected if they have co-authored a paper. The attributes of authors are terms that appear in the titles of papers authored by them[3]. In the DBLP dataset an attribute set defines a topic (i.e., set of terms that carry a specific meaning in the literature) and a dense subgraph is a community.

The DBLP dataset has 108,030 vertices, 276,658 edges and 23,285 attributes. Table 2 shows the top 10 attribute

---
[1]http://code.google.com/p/scpm/
[2]http://www.informatik.uni-trier.de/~ley/db
[3]Stemming and removal of stop words were applied.

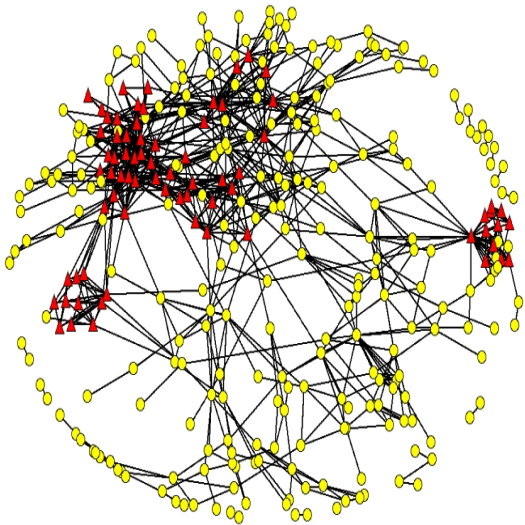
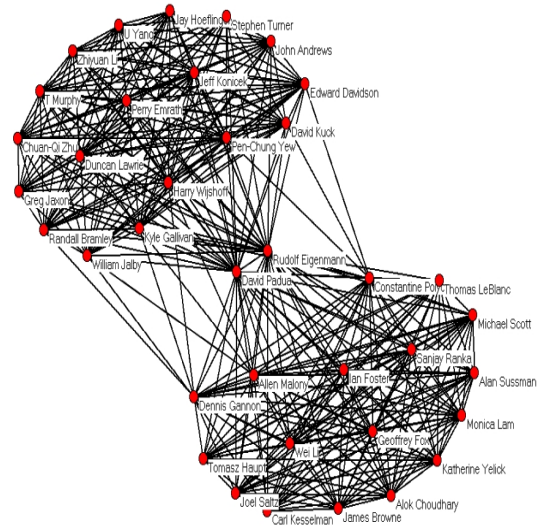

(a) Graph induced by $\{search, rank\}$  (b) Pattern induced by $\{perform, system\}$

Figure 3: Examples of results from the DBLP dataset

| $\sigma$ | | | | $\epsilon$ | | | | $\delta_{lb}$ | | | |
|---|---|---|---|---|---|---|---|---|---|---|---|
| S | $\sigma$ | $\epsilon$ | $\delta_{lb}$ | S | $\sigma$ | $\epsilon$ | $\delta_{lb}$ | S | $\sigma$ | $\epsilon$ | $\delta_{lb}$ |
| base system | 5492 | .04 | 14.0 | grid applic | 840 | .26 | 41577 | search rank | 420 | .19 | 635349 |
| base us | 5421 | .04 | 13.5 | grid servic | 599 | .23 | 154703 | perform file | 404 | .14 | 555067 |
| base model | 4852 | .03 | 13.3 | environ grid | 525 | .21 | 256793 | structur index | 404 | .14 | 555067 |
| model us | 4168 | .03 | 21.0 | queri xml | 615 | .21 | 123533 | search mine | 413 | .14 | 490932 |
| system us | 3989 | .05 | 36.8 | search web | 1031 | .20 | 13738 | us xml | 400 | .11 | 442638 |
| base network | 3774 | .05 | 41.8 | search rank | 420 | .19 | 635349 | search web data | 424 | .14 | 431589 |
| model system | 3460 | .02 | 21.7 | dynam simul | 469 | .19 | 383169 | base search analysi | 414 | .12 | 416385 |
| base data | 3452 | .07 | 71.6 | queri data | 1540 | .19 | 2758 | model internet | 401 | .10 | 406059 |
| base imag | 3424 | .02 | 17.6 | chip system | 702 | .19 | 63351 | process data databas | 416 | .12 | 405363 |
| imag us | 3345 | .02 | 19.6 | data stream | 1073 | .18 | 10653 | perform distribut parallel | 416 | .11 | 388818 |

Table 2: DBLP - Top support ($\sigma$), str. correlation ($\epsilon$), and normalized str. correlation ($\delta_{lb}$) attribute sets.

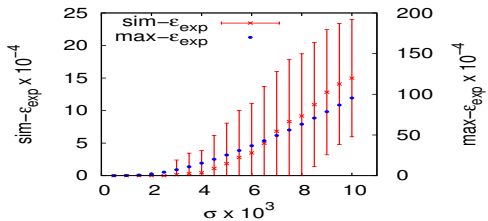

Figure 4: DBLP - Expected $\epsilon$ computed by the simulation (sim-$\epsilon_{exp}$) and analytical (max-$\epsilon_{exp}$) models.

sets w.r.t support ($\sigma$), structural correlation ($\epsilon$), and normalized structural correlation ($\delta_{lb}$). The minimum size ($min\_size$) and density ($\gamma_{min}$) parameters were set to 10 and 0.5, respectively. The minimum support threshold ($\sigma_{min}$) was set to 400 and we considered only attribute sets of size at least 2. The parameters used in our case studies were selected empirically.

Top-$\sigma$ attribute sets present a low correlation with the formation of dense subgraphs in the DBLP dataset. Such terms are popular in paper titles, but do not carry much knowledge regarding the formation of research communities. On the other hand, top-$\epsilon$ structural correlation may be more easily associated to known topics in computer science. The attribute set $\{grid, applic\}$ has the highest structural correlation (0.26), i.e., 26% of the authors that have the keywords "grid" and "applic" are inside a community of researchers of size at least 10 where each of them have collaborated with half of the other members. It is interesting to point out that the graph induced by $\{grid, applic\}$ has more vertices in dense subgraphs than the graph induced by $\{base, system\}$, though $\{base, system\}$ is more than 6 times more frequent than $\{grid, applic\}$. In general, high support attribute sets do not present high structural correlation.

Figure 4, shows the expected structural correlation for different support values in the DBLP dataset. The input parameters are the same as those used to generate the results shown in Table 2. For the simulation model, we executed 1000 simulations for each support value and show also the standard deviation of the expected structural correlation estimated. The analytical upper bound is not tight w.r.t. the simulation results, but presents a similar growth, which shows that it enables accurate comparisons between the structural correlation of attribute sets.

Based on the proposed analytical model, the third column of Table 2 shows the top attribute sets in terms of analytical normalized structural correlation ($\delta_{lb}$). The attribute set

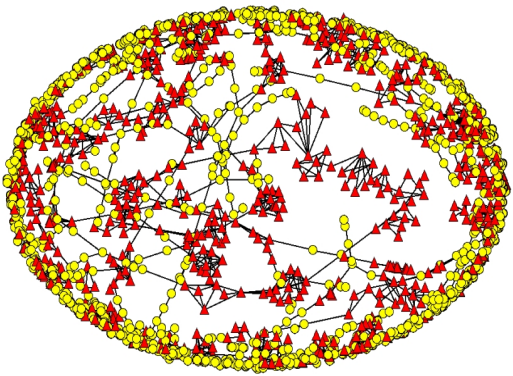

(a) Graph induced by {S Stevens,Wilco}

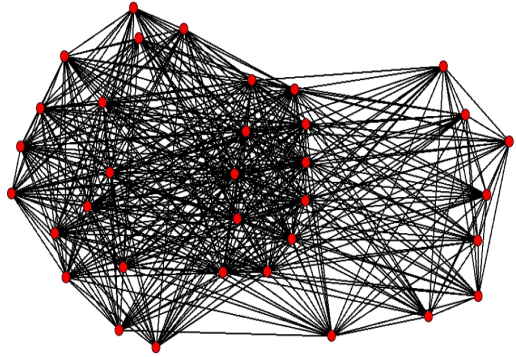

(b) Pattern induced by {Van Morrison}

Figure 5: Examples of results from the LastFm dataset

| $\sigma$ | | | | $\epsilon$ | | | | $\delta_{lb}$ | | | |
|---|---|---|---|---|---|---|---|---|---|---|---|
| S | $\sigma$ | $\epsilon$ | $\delta_{lb}$ | S | $\sigma$ | $\epsilon$ | $\delta_{lb}$ | S | $\sigma$ | $\epsilon$ | $\delta_{lb}$ |
| Radiohead | 121892 | .11 | .37 | Radiohead | 121892 | .11 | .37 | S Stevens,Wilco | 28798 | .04 | 1.14 |
| Coldplay | 118053 | .09 | .33 | Coldplay | 118053 | .09 | .33 | S Stevens,Of Montreal | 28621 | .04 | 1.13 |
| Beatles | 109037 | .09 | .36 | Beatles | 109037 | .09 | .36 | Beirut | 27605 | .04 | 1.11 |
| R Peppers | 105984 | .09 | .35 | R Peppers | 105984 | .09 | .35 | S Stevens,Decemberists,Beatles | 27415 | .04 | 1.11 |
| Nirvana | 100604 | .07 | .31 | Metallica | 83587 | .08 | .41 | N Hotel,S Stevens | 29260 | .04 | 1.10 |
| T Killers | 96305 | .07 | .32 | DC for Cutie | 82025 | .07 | .41 | S Stevens,F Lips,Beatles | 27571 | .04 | 1.09 |
| Muse | 94382 | .07 | .33 | Beck | 83360 | .07 | .40 | A Collective | 33555 | .05 | 1.09 |
| Oasis | 87875 | .06 | .30 | Muse | 94382 | .07 | .33 | BS Scene,NM Hotel | 27308 | .04 | 1.09 |
| F Fighters | 87001 | .06 | .33 | Nirvana | 100604 | .07 | .31 | Radiohead,Spoon,S Stevens | 27113 | .04 | 1.06 |
| P Floyd | 86807 | .07 | .34 | The Shins | 68480 | .07 | .50 | N Hotel,Radiohead,Beatles | 28776 | .04 | 1.04 |

Table 3: LastFm - Top support ($\sigma$), str. correlation ($\epsilon$), and normalized str. correlation ($\delta_{lb}$) attribute sets.

{$search, rank$} has the highest normalized structural correlation (635,349), i.e., the structural correlation is 635,349 times the upper bound on its expected structural correlation given by the analytical model. Figure 3(a) presents the graph induced by {$search, rank$}. Vertices contained in a dense subgraph are indicated. Dense subgraphs cover the densest components of the induced graph. In general, top-$\sigma$ attribute sets have low $\delta_{lb}$ when compared to the top-$\delta_{lb}$ attribute sets. Moreover, high values of $\epsilon$ do not necessarily lead to high values of $\delta_{lb}$. Figure 3(b) shows the largest structural correlation pattern in terms of number of vertices from DBLP, which represents two important interconnected research groups on high performance systems.

### 4.1.2 LastFm

LastFm[4] is an online social music network. We use a sample of the LastFm users crawled through an API provided by LastFm. In the LastFm network, vertices represent users and edges represent friendships. The attributes of a vertex are the artists the respective user has listened to. An attribute set in the LastFm dataset represents, in a more general interpretation, a musical taste (i.e., set of artists) and a dense subgraph is a community.

The LastFm dataset contains 272,412 vertices, 350,239 edges, and 3,929,101 attributes. Table 3 shows the top 10 attribute sets in terms of support ($\sigma$), structural correlation ($\epsilon$) and normalized structural correlation ($\delta_{lb}$) discovered from LastFm. The minimum size ($min\_size$) and density ($\gamma_{min}$) parameters were set to 5 and 0.5, respectively. The minimum support threshold ($\sigma_{min}$) was set to 27,000.

[4] http://www.last.fm

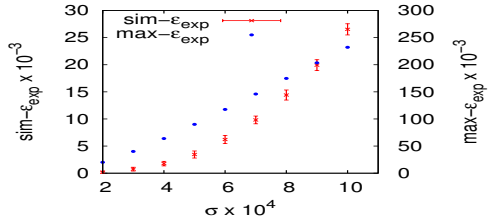

Figure 7: LastFm - Expected $\epsilon$ computed by the simulation (sim-$\epsilon_{exp}$) and analytical (max-$\epsilon_{exp}$) models.

In general, the top-$\epsilon$ attribute sets are the most frequent ones. However, such attribute sets present low normalized structural correlation. In other words, although these attributes are frequent and have several vertices covered by communities, this coverage is not much higher than expected. Considering the normalized structural correlation, which takes into account the expected structural correlation of an attribute set, the top patterns change significantly. Figure 7 shows the expected structural correlation for support values varying from 20,000 to 100,000. Each simulation-based expected structural correlation value corresponds to an average of 100 simulations. The top $\delta_{lb}$ attribute set {S Stevens,Wilco} includes the American singer and songwriter Sufjan Stevens and the American band Wilco.

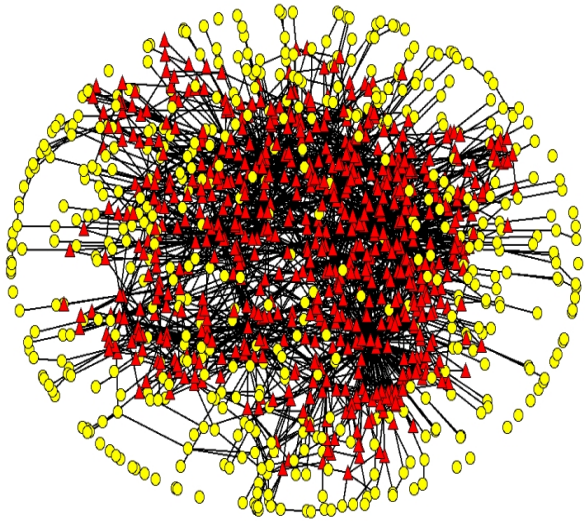
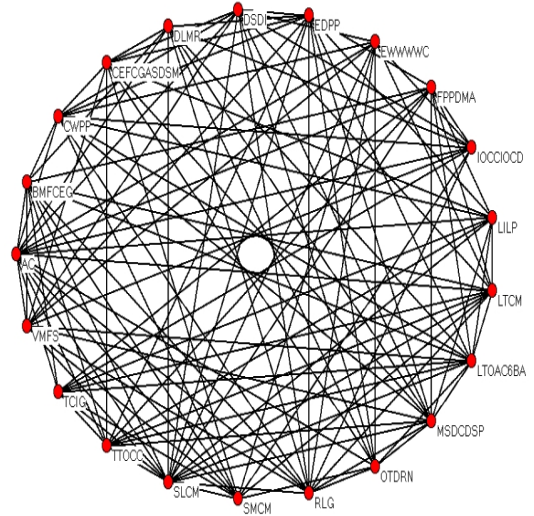

(a) Graph induced by {node,wireless}     (b) Pattern induced by {perform,system}

Figure 6: Examples of results from the CiteSeer dataset

| $\sigma$ | | | | $\epsilon$ | | | | $\delta_{lb}$ | | | |
|---|---|---|---|---|---|---|---|---|---|---|---|
| S | $\sigma$ | $\epsilon$ | $\delta_{lb}$ | S | $\sigma$ | $\epsilon$ | $\delta_{lb}$ | S | $\sigma$ | $\epsilon$ | $\delta_{lb}$ |
| system paper | 57906 | .16 | .77 | network sensor | 3276 | .47 | 108.7 | node wireless | 2086 | .35 | 164.4 |
| base paper | 56566 | .10 | .47 | network hoc | 2744 | .47 | 141.2 | protocol rout | 2134 | .35 | 157.6 |
| paper result | 47516 | .08 | .45 | ad network hoc | 2725 | .44 | 134.6 | memori cach | 2150 | .32 | 143.8 |
| paper model | 43929 | .09 | .59 | network rout | 5084 | .41 | 48.0 | network hoc | 2744 | .47 | 141.2 |
| us paper | 43573 | .05 | .32 | network wireless | 5242 | .40 | 44.7 | protocol wireless | 2048 | .29 | 138.7 |
| system base | 42079 | .09 | .63 | node wireless | 2086 | .35 | 164.4 | ad network hoc | 2725 | .44 | 134.7 |
| approach paper | 38690 | .05 | .40 | protocol rout | 2134 | .35 | 157.6 | network node rout | 2075 | .25 | 118.3 |
| perform paper | 37349 | .13 | 1.04 | ad network | 3563 | .34 | 69.3 | optim queri | 2094 | .26 | 118.2 |
| paper propos | 37243 | .06 | .46 | program logic | 5895 | .33 | 31.2 | perform instruct | 2111 | .25 | 115.95 |
| paper algorithm | 37027 | .12 | .95 | memori cach | 2150 | .32 | 143.8 | paper ad network | 2081 | .23 | 108.86 |

Table 4: CiteSeer - Top support ($\sigma$), str. correlation ($\epsilon$), and normalized str. correlation ($\delta_{lb}$) attribute sets.

Figure 5(a) shows the graph induced by the attribute set {S Stevens,Wilco}. For clarity, we removed vertices with degree lower than 2. By visualizing vertices inside and outside structural correlation patterns, we can understand how the structural correlation captures the relationship between attributes and dense subgraphs. The largest structural correlation pattern found is presented in Figure 5(b). It represents a community of 34 users who have listened to the Northern Irish singer and songwriter Van Morrison. Vertex identifiers are not shown due to privacy issues.

### 4.1.3 CiteSeer

CiteSeerX[5] is a scientific literature digital library and search engine. We built a citation graph from CiteSeerX as of March of 2010. In the CiteSeer graph, papers are represented by vertices and citations by undirected edges. Each paper has as attributes terms extracted from its abstract[6]. Attribute sets represent topics and dense subgraphs define groups of related work in the CiteSeer graph.

The CiteSeer dataset has 294,104 vertices, 782,147 edges, and 206,430 attributes. The parameters setting applied in this case study is $\sigma_{min} = 2000$, $min\_size = 5$, and $\gamma_{min} = 0.5$. Table 4 shows the top structural correlation attribute sets w.r.t. $\sigma$, $\epsilon$, and $\delta_{lb}$ discovered. Top-$\sigma$ attribute sets present low structural correlation and normalized structural correlation when compared to the top-$\epsilon$ and top-$\delta_{lb}$ attribute sets, respectively. Moreover, similar to the DBLP dataset, while the top-$\sigma$ attribute sets from the CiteSeer dataset are generic terms, the top-$\epsilon$ and top-$\delta_{lb}$ attribute sets may be easily associated to known research topics (e.g, computer networks, query optimization).

[5]http://citeseerx.ist.psu.edu
[6]Stemming and stop words removal were applied.

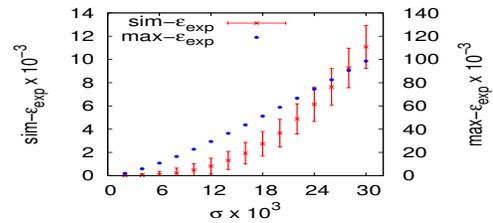

Figure 9: CiteSeer - Expected $\epsilon$ for the simulation (sim-$\epsilon_{exp}$) and analytical (max-$\epsilon_{exp}$) models.

Figure 9 shows the expected structural correlation for different support values in CiteSeer. The attribute set {*node, wireless*} has the highest normalized structural correlation ($\delta_{lb}$ =164.40). Figure 6(a) shows the graph induced by the attribute set $\{node, wireless\}$ in CiteSeer. Figure 6(b) presents the largest structural correlation pattern discovered in the CiteSeer dataset. Vertex labels are the initials of paper titles. The papers included in the pattern cover topics such as caching, memory management, computer networks, processor design, and instruction level optimization (e.g., Attribute Caches, Systems for Late Code Modification, Limits of Instruction Level Parallelism, Link-time Optimization of Address Calculation on a 64-bit Architecture). We do not show the full list of paper titles due to space limitations.

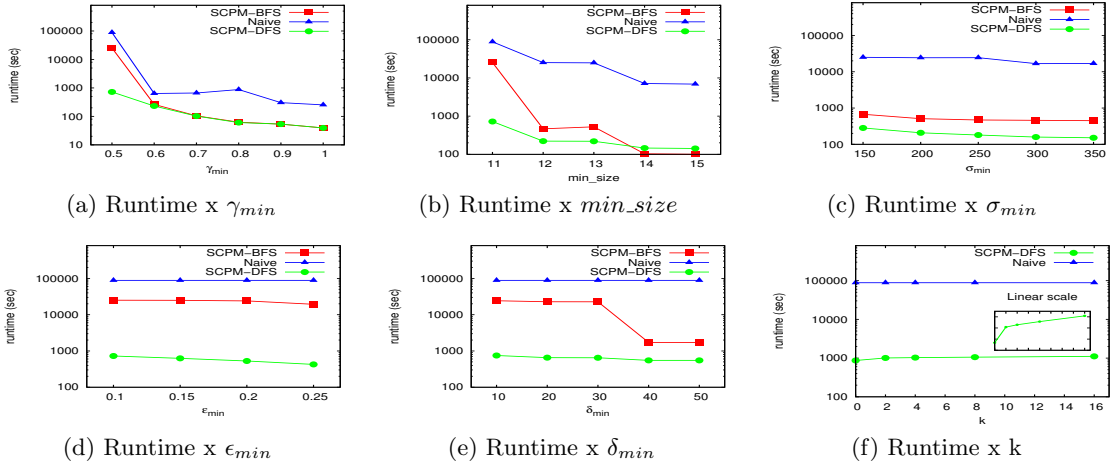

Figure 8: Performance evaluation

## 4.2 Performance Evaluation

This section evaluates the performance of the structural correlation pattern mining algorithms. The dataset used is a smaller version of the DBLP dataset (SmallDBLP), which has 32,908 vertices, 82,376 edges, and 11,192 attributes.

The **SCPM-BFS** and **SCPM-DFS** are versions of the SCPM algorithm using the BFS and DFS strategy, respectively. The **Naive** algorithm enumerates the complete set of quasi-cliques from the induced graphs, as described in Section 3.1. We vary each parameter of the algorithms keeping the others constant. Default values for $\gamma_{min}$, $min\_size$, and $\sigma_{min}$ are 0.5, 11, and 100. Moreover $\epsilon_{min}$, $\delta_{min}$, and k are set to 0.1, 1, and 5, respectively, unless stated otherwise.

Figures 8(a), 8(b), and 8(c) show the runtime of the algorithms varying the values of $\gamma_{min}$, $min\_size$, and $\sigma_{min}$, respectively. In general, **SCPM-DFS** achieves the best results, being up to 3 orders of magnitude faster than the **Naive** algorithm. Moreover, **SCPM-BFS** performs better than the **Naive** algorithm in all the experiments.

In terms of the $\epsilon_{min}$ (Figure 8(d)) and $\delta_{min}$ (Figure 8(e)) parameters, both the **SCPM-BFS** and **SCPM-DFS** apply the pruning techniques described in Section 3.2.1. Based on the results shown in Figures 8(d) and 8(e), we can notice that such techniques lead to significant performance gains when the values of $\epsilon_{min}$ and $\delta_{min}$ are increased.

In Figure 8(f), we show the runtime of **SCPM-DFS** and the **Naive** algorithm for different values of k. The results of **SCPM-BFS** are omitted because both **SCPM-BFS** and **SCPM-DFS** algorithms apply the same strategy for identifying the top-k structural correlation patterns (see Section 3.2.3). The inset also shows the execution time of **SCPM-DFS** using a linear scale for the y-axis, to more clearly see the effect of k on the runtime. The results show that for low values of top k, **SCPM-DFS** is able to achieve low running times, outperforming the **Naive** algorithm significantly.

## 4.3 Parameter Sensitivity and Setting

We now assess how different input parameters affect the output of structural correlation pattern mining. Our objective is to provide guidelines for setting the parameters of SCPM. Figure 10 shows the average structural correlation and normalized structural correlation of the complete output (global) and the top-10% attribute sets from the SmallDBLP dataset varying the $\gamma_{min}$, $min\_size$, and $\sigma_{min}$ parameters. Default values for $\gamma_{min}$, $min\_size$, and $\sigma_{min}$ are 0.5, 10 and 100. The results show that more restrictive quasi-clique parameters (i.e., high values of $\gamma_{min}$ and $min\_size$) reduce the average $\epsilon$ but may increase $\delta$, since dense subgraphs become less expected. Moreover, high values of $\sigma_{min}$ are related to high values of structural correlation $\epsilon$. However, such attribute sets also present high values of $\epsilon_{exp}$, leading to low values of normalized structural correlation $\delta$.

SCPM is an exploratory pattern mining method, and thus reasonable values for the different parameters can be obtained by searching the parameter space. The minimum density parameter, $\gamma_{min}$, and the minimum quasi-clique size, $min\_size$, will depend on the application. For $\sigma_{min}$, a useful guideline is to select values that produce a significant expected structural correlation. Infrequent attribute sets may not be expected to induce any dense subgraph. The other parameters ($\epsilon_{min}$, $\delta_{min}$, and k) have as objectives to speedup the algorithm and must be set according to the available computational resources and time.

## 5. CONCLUSIONS

In this paper, we studied the problem of correlating vertex attributes and dense subgraphs in large attributed graphs. The concept of structural correlation, which measures how an attribute set induces dense subgraphs in an attributed graph was proposed. We also presented normalization approaches that compare the structural correlation of a given attribute set against its expected value, which provides a measure of the statistical significance for the structural correlation. In order to enable the analysis of large databases, we introduced search and pruning strategies for structural correlation pattern mining. We also proposed an algorithm for the identification of the top structural correlation patterns, which are the largest and densest subgraphs induced by a given set of attributes. The patterns and algorithms proposed were applied to three real datasets. The attribute sets and patterns found represent relevant knowledge in terms of the correlation between attributes and dense subgraphs.

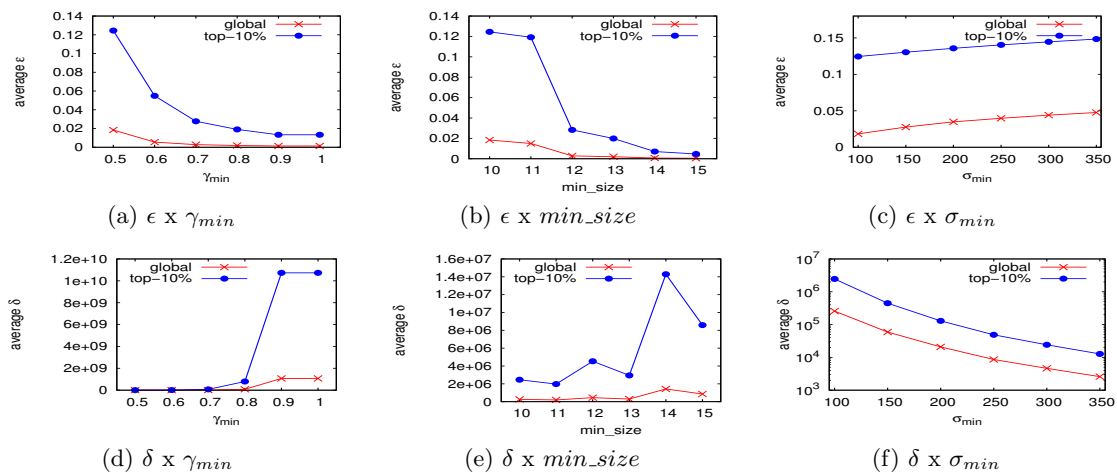

Figure 10: Parameter sensitivity

**Acknowledgements:** This work was supported by CNPQ, CAPES, Fapemig, FINEP, InWeb, NSF award EMT-0829835, and NIH award 1R01EB0080161. We would like to thank Alberto Laender and Loïc Cerf for their comments.